\documentclass[3p,twocolumn]{elsarticle}

\usepackage{verbatim}
\usepackage{geometry}
\usepackage[T1]{fontenc}
\usepackage[utf8]{inputenc}
\usepackage[american]{babel}
\usepackage{epsfig}
\usepackage{graphicx,subcaption,caption}
\usepackage{booktabs}
\usepackage{multirow}
\usepackage{dcolumn}
\usepackage{amsmath}
\usepackage{mathtools}
\usepackage{amsfonts}
\usepackage{amssymb}
\usepackage{epstopdf}
\usepackage{bm}
\usepackage{siunitx}
\usepackage{braket}
\usepackage{enumitem}
\usepackage{hyperref}
\usepackage{soul}
\usepackage[table]{xcolor}
\usepackage{color}
\usepackage{transparent}
\usepackage{pifont}
\usepackage{orcidlink}
\biboptions{sort&compress}
\usepackage[normalem]{ulem}
\usepackage[capitalise]{cleveref}
\crefname{subsection}{Subsection}{Subsections}
\Crefname{subsection}{Subsection}{Subsections}

\definecolor{navyblue}{rgb}{0.0, 0.0, 0.5}
\definecolor{royalblue}{rgb}{0.25, 0.41, 0.88}
\definecolor{cadmiumgreen}{rgb}{0.0, 0.42, 0.24}
\definecolor{blue-violet}{rgb}{0.54, 0.17, 0.89}
\definecolor{darkviolet}{rgb}{0.58, 0.0, 0.83}
\definecolor{orange(colorwheel)}{rgb}{1.0, 0.5, 0.0}

\usepackage{hyperref}
\hypersetup{
    colorlinks=true, % false: boxed links; true: colored links
    linkcolor=darkviolet, % color of internal links (change box color with linkbordercolor)
    citecolor=darkviolet}

\usepackage{booktabs}
\usepackage{multirow}
\usepackage{dcolumn}
\usepackage{colortbl}
\usepackage{url}
\usepackage{hyperref}

\begin{document}

\begin{frontmatter}

\title{Present Day Cosmic Acceleration from SDSS and DESI BAO:\\ A Call for Finer Tomography of the DESI Bright Galaxy Survey}

\author[1]{Anna Chiara Ferri}
\author[2]{Ruchika}
\author[1]{Alessandro Melchiorri}

\address[1]{Physics Department and INFN, Universit\`a di Roma ``La Sapienza'', P.le Aldo Moro 2, 00185, Rome, Italy}
\address[2]{Departamento de Física Fundamental and IUFFyM, Universidad de Salamanca, E-37008 Salamanca, Spain}

\begin{abstract}
The DESI collaboration's Data Release~2 (DR2) provides baryon acoustic oscillation (BAO) measurements from over 14 million galaxies and quasars, and a joint analysis of DESI BAO, CMB, and Type~Ia Supernovae reveals a preference for time-evolving dark energy. We quantify this preference relative to SDSS BAO and report three key results. First, DESI+Planck favors a higher $w_0 = -0.41^{+0.21}_{-0.22}$ than SDSS+Planck ($w_0 = -0.71^{+0.19}_{-0.18}$). Second, DESI+Planck prefers a deceleration parameter whose median lies on the decelerating side ($q_0 = 0.10^{+0.21}_{-0.23}$, consistent with $q_0 = 0$ at $1\sigma$), while SDSS+Planck prefers a negative value ($q_0 =-0.22^{+0.20}_{-0.21}$) indicating accelerated expansion. Third, we argue that this discrepancy arises from the difference in the lowest effective redshift probed by each survey: $z_{\rm eff} \approx 0.295$ for DESI versus $z_{\rm eff} \approx 0.15$ for SDSS.\\
As present-day quantities, $w_0$ and $q_0$ are sensitive to the lowest probed redshift: data near $z = 0$ constrain them directly, whereas higher-redshift data rely on extrapolating the dark energy parametrization (here CPL). Reaching $z_{\rm eff} \approx 0.15$, SDSS constrains $w_0$ and $q_0$ in a data-driven way, finding consistency with $w_0 = -1$ and acceleration. Limited to $z_{\rm eff} \gtrsim 0.295$, DESI relies more on extrapolation, driving $q_0$ positive and $w_0$ well above $-1$. Adding the Pantheon+ supernova sample restores low-redshift information, returning $q_0$ to negative values and reducing tension with $\Lambda\text{CDM}$. We therefore propose that the apparent DESI preference for a non-accelerating present epoch in the BAO+CMB combination reflects redshift sampling rather than new physics, and suggest future DESI analyses adopt finer tomographic binning of the Bright Galaxy Survey sample to access lower mean redshifts and test this conclusion.
\end{abstract}


\end{frontmatter}

%==================================================================
\section{Introduction}
%==================================================================

The Dark Energy Spectroscopic Instrument (DESI)~\cite{DESI:2016fyo,DESI:2016igz,DESI:2022gle} has, over its first two data
releases, transformed the precision landscape of baryon acoustic oscillation
(BAO) cosmology~\cite{Eisenstein:2005su,Cole:2005sx}. The first data release (DR1) provided BAO measurements in
seven redshift bins using galaxy, quasar, and Lyman-$\alpha$ forest tracers
across $0.1 < z < 4.2$, expressed in terms of the transverse comoving distance
$D_M/r_d$, the Hubble distance $D_H/r_d$, or their angle-averaged combination
$D_V/r_d$, each normalized to the comoving sound horizon at the end of the
baryon drag epoch~\cite{DESI:2024mwx,DESI:2024uvr,DESI:2024lzq}. While the DESI BAO data alone are consistent
with the standard $\Lambda$CDM model~\cite{Planck:2018vyg}, their combination with external datasets
yields results in tension with the concordance scenario~\cite{DESI:2024mwx,DESI:2024kob}. Allowing for a
time-varying dark energy equation of state parametrized by
$w(a) = w_0 + w_a(1-a)$~\cite{Chevallier:2000qy,Linder:2002et}, the combination of DESI with cosmic microwave
background (CMB)~\cite{Planck:2018vyg} or Type Ia supernovae (SNe Ia) measurements individually
favors $w_0 > -1$ and $w_a < 0$. In the first data release (DR1), the joint analysis of DESI, CMB, and SN~Ia
probes is discrepant with $\Lambda$CDM at the $2.5\sigma$, $3.5\sigma$, and
$3.9\sigma$ levels when using the Pantheon+~\cite{Scolnic:2021amr,Brout:2022vxf}, Union3~\cite{Rubin:2023ovl},
or Dark Energy Survey (DES) Y5~\cite{DES:2024tys} supernova compilations,
respectively.

The second data release (DR2) substantially sharpens this picture~\cite{DESI:2025zgx,DESI:2025fii}. Since DR1 is
a subset of DR2 and the reduction and analysis pipelines are nearly identical,
the two can be compared directly; DR2 delivers a $30$--$50\%$ improvement in
statistical precision while remaining highly consistent with DR1 (a
Kolmogorov--Smirnov test on the differences yields a $p$-value of $0.40$).
Internally, the DR2 BAO measurements across redshift bins show no inconsistency
within $\Lambda$CDM, with the largest difference in recovered parameters being
$1.8\sigma$ (between LRG1 and LRG3+ELG1). With more than 14 million galaxies and quasars, DR2 thus reinforces
rather than dilutes the DR1 hint of dynamical dark energy~\cite{DESI:2025zgx}: the joint DESI~DR2, CMB, and SN~Ia analyses disfavor $\Lambda$CDM at the $2.8\sigma$, $3.8\sigma$, and $4.2\sigma$ levels with Pantheon+, Union3, and DES~Y5, respectively~\cite{DESI:2025zgx}.

The region of sky and the redshift range observed by DESI partially overlap
with those of the earlier BOSS~\cite{BOSS:2012dmf,BOSS:2016wmc} and eBOSS~\cite{SDSS:2015jyv} programs of
SDSS~\cite{SDSS:2000hjo}, whose final BAO results were presented in~\cite{eBOSS:2020yzd,Bautista:2020ahg,deMattia:2020fkb,Tamone:2020qrl,Hou:2020rse,Neveux:2020voa,duMasDesBourboux:2020pck} (see also~\cite{Ruchika:2024lgi,CosmoVerseNetwork:2025alb,Dutta:2019pio,Capozziello:2018jya} for related model-independent analyses of low redshift data).
This overlap motivates a direct comparison between the two surveys. In DR1, DESI
reported systematically larger observed BAO scales than the Planck 2018
$\Lambda$CDM prediction~\cite{Planck:2018vyg} at $z < 0.8$~\cite{Chudaykin:2024gol,DESI:2025zgx,Colgain:2024xqj,Colgain:2024mtg,Mukherjee:2024ryz}, and correspondingly smaller distances. The
most prominent feature was an approximately $3\sigma$ difference between the
DESI measurement of $D_M/r_d$ in the LRG2 bin at $z_{\rm eff} = 0.71$ and the
corresponding SDSS result at $z = 0.7$, with the LRG1 bin at $z_{\rm eff} =
0.51$ also contributing. Above $z = 0.8$ the DESI and SDSS measurements agree
well, so the difference between the two BAO datasets originates almost entirely
from $z < 0.8$. In DR2, the new measurement in this bin shifts to lie between
the DR1 DESI and SDSS values in the $D_M/r_d$--$D_H/r_d$ plane, partially easing
the original discrepancy~\cite{DESI:2025zgx}.

These observations have driven a body of literature focused on the low redshift
behavior of DESI~\cite{Cortes:2024lgw,Carloni:2024zpl,Giare:2024gpk}. The standard consistency check has been to replace the DESI
data below $z \simeq 0.6$--$0.8$ with the corresponding SDSS measurements and to
re-examine the evidence for evolving dark energy~\cite{Chudaykin:2024gol,DES:2026jmi,eBOSS:2020yzd,Wang:2025bkk,Wang:2024rjd}. The DESI collaboration itself
performed such a test, finding good agreement with the baseline analysis while
the $w_0$--$w_a$ constraints shifted closer to $\Lambda$CDM and the tension
marginally decreased to $2.1\sigma$ with Pantheon+. Subsequent
studies~\cite{Sapone:2024ltl,Efstathiou:2024xcq} have likewise identified the LRG1 ($z_{\rm eff} =
0.51$) and LRG2 ($z_{\rm eff} = 0.71$) bins as the primary drivers of the DESI
preference for the $w_0w_a$ model relative to $\Lambda$CDM.

The remainder of this paper is organized as follows. In
Section~\ref{sec:expansion} we set out the background expansion history and the
dark energy parametrizations (CPL, $w$CDM, and the massive neutrino extensions).
In Section~\ref{sec:3} we define the deceleration parameter $q(z)$, derive $q_0$
in terms of $w_0$ and $\Omega_{\rm DE}$, and state the condition for present day
acceleration. In Section~\ref{sec:Data} we describe the DESI and SDSS BAO
datasets and the external CMB and SN~Ia information. We present our results and
the comparison between the two surveys in Section~\ref{sec:Results}, and discuss
and conclude in Sections~\ref{sec:Discussion} and~\ref{sec:Conclusions}.

%==================================================================
\section{Expansion History}\label{sec:expansion}
%==================================================================

The expansion history of the Universe is fundamentally described by the Hubble parameter $H(z) \equiv \dot{a}/a$, which represents the rate of change of the scale factor $a(t)$. In a spatially flat Friedmann--Lema\^itre--Robertson--Walker (FLRW) metric, the late-time dynamics is governed by the acceleration equation derived from the Friedmann equations~\cite{Carroll2001,Frieman2008,Copeland2006,CaldwellKamionkowski2009,Amendola:2012ys}:
\begin{equation}
\frac{\ddot a}{a} = -\frac{4\pi G}{3} \left(\rho + 3p \right),
\label{eq:acc}
\end{equation}
where $\rho$ is the total energy density and $p$ the total pressure of the cosmic fluid. For a Universe composed of pressureless matter ($\rho_m$) and a dark energy component ($\rho_{\rm DE}$), the total density and pressure are
\begin{equation}
\rho = \rho_m + \rho_{\rm DE}, \qquad p = p_m + p_{\rm DE}.
\end{equation}
Since matter is pressureless ($p_m = 0$) and dark energy follows the equation of state $p_{\rm DE} = w(z)\,\rho_{\rm DE}$, the acceleration equation~(\ref{eq:acc}) becomes
\begin{equation}
\frac{\ddot a}{a} = -\frac{4\pi G}{3} \left[ \rho_m + \rho_{\rm DE}\bigl(1+3w(z)\bigr) \right],
\end{equation}
which makes explicit that cosmic acceleration ($\ddot{a} > 0$) requires $\rho + 3p < 0$, a condition physically achievable only if the dark energy component supplies a sufficiently negative pressure~\cite{Carroll2001,Frieman2008,Copeland2006}.

In terms of the redshift $z$, the Hubble parameter evolves as
\begin{equation}
H^2(z) = H_0^2 \left[ \Omega_m (1+z)^3 + \Omega_{\rm DE}\, f_{\rm DE}(z) \right],
\label{eq:Hz}
\end{equation}
where $H_0$ is the Hubble constant, $\Omega_i$ are the present day fractional energy densities, and $f_{\rm DE}(z)$ encodes the time evolution of the dark energy density~\cite{Carroll2001,Frieman2008,Copeland2006,Amendola:2012ys}. The functional form of $f_{\rm DE}(z)$ depends on the chosen parametrization of $w(z)$, and several physically motivated options can be considered~\cite{Chevallier:2000qy,Linder:2002et,Copeland2006}.

\textit{CPL parametrization:} In the Chevallier--Polarski--Linder (CPL) parametrization, also known as the $w_0$--$w_a$ parametrization~\cite{Chevallier:2000qy,Linder:2002et,Caldwell:2005tm}, the equation of state is expanded linearly in the scale factor,
\begin{equation}
w(a) = w_0 + w_a (1-a),
\end{equation}
so that the dark energy evolution factor reads
\begin{equation}
f_{\rm DE}(z) = (1+z)^{3(1+w_0+w_a)} \exp\!\left[-3 w_a \frac{z}{1+z}\right].
\label{eq:fdeCPL}
\end{equation}
This is the parametrization adopted by the DESI collaboration~\cite{DESI:2024mwx,DESI:2025zgx} and the one in which we will report our main results, since it allows a direct comparison with the SDSS analysis (see Fig.~12 of Ref.~\cite{eBOSS:2020yzd}) and a transparent decomposition of $q_0$ in terms of $w_0$ and $\Omega_m$~\cite{Sahni:2008xx,Farooq:2016zwm,Wang:2018fyk}.

\textit{wCDM parametrization:}
The $w$CDM model is the minimal extension of $\Lambda$CDM in which the dark energy equation of state $w$ is constant but free~\cite{Carroll2001,Frieman2008,Copeland2006,Caldwell:2005tm,Amendola:2012ys}; it is recovered from CPL by setting $w_a = 0$~\cite{Chevallier:2000qy,Linder:2002et}. In this case,
\begin{equation}
f_{\rm DE}(z) = (1+z)^{3(1+w_0)}.
\label{eq:fdewCDM}
\end{equation}

\textit{Massive neutrino extensions:} In the standard cosmological model the total neutrino mass $\sum m_\nu$ enters as a separate component whose effect on the background expansion is partly degenerate with that of dark energy at low redshift~\cite{Lesgourgues:2006nd,Hannestad:2005gj,Vagnozzi:2017ovm,Lattanzi:2017ubd}. When neutrinos are non-relativistic at $z=0$, equation~(\ref{eq:Hz}) generalizes to
\begin{equation}
H^2(z) = H_0^2 \left[ \Omega_m (1+z)^3 + \Omega_\nu(z) + \Omega_{\rm DE}\, f_{\rm DE}(z) \right],
\end{equation}
where $\Omega_\nu(z)$ accounts for the redshift-dependent transition of neutrinos from the relativistic to the non-relativistic regime~\cite{Lesgourgues:2006nd,Lattanzi:2017ubd}. Two cases are of interest here:
\begin{itemize}
    \item \textbf{$\Lambda$CDM$+\sum m_\nu$} (denoted $\nu$CDM in the following), where the dark energy sector is still a cosmological constant and $f_{\rm DE}(z) = 1$;
    \item \textbf{$w$CDM$+\sum m_\nu$}, where $f_{\rm DE}(z)$ takes the constant-$w$ form of equation~(\ref{eq:fdewCDM}).
\end{itemize}
These extensions are particularly relevant in light of the well known degeneracy between $\sum m_\nu$ and $w$ in CMB+BAO analyses, as a non-zero neutrino mass can partially mimic a dark energy component with $w \neq -1$ through its impact on the late-time expansion~\cite{Hannestad:2005gj,Lesgourgues:2006nd,Planck:2018vyg,Vagnozzi:2017ovm,Lattanzi:2017ubd}. They are included in our analysis as comparison cases. We plan to extend this analysis to other dynamical dark energy parametrizations in future work~\cite{Doran:2006kp,Zhao:2017cud,Wang:2018fyk,Sahni:2014ooa}.

%==================================================================
\section{Deceleration Parameter}\label{sec:3}
%==================================================================

A direct diagnostic of the expansion dynamics is the dimensionless deceleration parameter $q(z)$, defined as~\cite{Carroll2001,Frieman2008,Sahni:2008xx}
\begin{equation}
q(z) \equiv -\frac{\ddot{a}}{a H^2} = -1 + \frac{(1+z)}{H(z)}\frac{dH(z)}{dz}.
\end{equation}
Using the Friedmann equation $H^2 = \tfrac{8\pi G}{3}(\rho_m + \rho_{\rm DE})$ and the density parameters $\Omega_i = \rho_i / \rho_{\rm crit}$, the present day deceleration parameter takes the form
\begin{equation}
q_0 = \tfrac{1}{2}\Omega_m + \tfrac{1}{2}\bigl(1+3w_0\bigr)\Omega_{\rm DE}.
\end{equation}
Under the assumption of spatial flatness ($\Omega_m + \Omega_{\rm DE} = 1$), this simplifies to
\begin{equation}
q_0 = \tfrac{1}{2}(1-\Omega_{\rm DE}) + \tfrac{1}{2}\bigl(1+3w_0\bigr)\Omega_{\rm DE}.
\label{eq:q0flat}
\end{equation}
In the CPL parametrization $q_0$ depends only on $w_0$ and not on $w_a$, since $w_a$ controls the \emph{rate of change} of $w(z)$ and not its present day value~\cite{Chevallier:2000qy,Linder:2002et,Caldwell:2005tm,Sahni:2008xx}. This is precisely why the DESI vs.\ SDSS comparison at $z=0$ is so sensitive to how each dataset constrains $w_0$: any observation that effectively shifts $w_0$ by an amount $\Delta w_0$ produces a shift in $q_0$ of roughly $\tfrac{3}{2}\Omega_{\rm DE}\,\Delta w_0 \approx 1.0\,\Delta w_0$ for $\Omega_{\rm DE}\simeq 0.7$~\cite{DESI:2025zgx,eBOSS:2020yzd,Sahni:2014ooa,Farooq:2016zwm}.

\subsubsection*{Condition for cosmic acceleration}

Cosmic acceleration occurs when $q(z) < 0$~\cite{Carroll2001,Frieman2008,Moresco:2016mzx}. Setting $q_0 = 0$ in equation~(\ref{eq:q0flat}) yields the transition boundary between accelerating and decelerating present day cosmologies,
\begin{equation}
w_0 = -\frac{1}{3\,\Omega_{\rm DE}},
\end{equation}
which defines a curve in the $(w_0, \Omega_{\rm DE})$ plane separating the two regimes. For the fiducial value $\Omega_{\rm DE}\simeq 0.7$, this corresponds to $w_0 \simeq -0.48$: any equation of state less negative than this threshold is incompatible with a presently accelerating Universe, regardless of how negative $w$ may have been in the past~\cite{Carroll2001,Copeland2006,Caldwell:2005tm,Sahni:2008xx}.

A complementary diagnostic is provided by the \emph{transition redshift} $z_{\rm crit}$, defined implicitly by $q(z_{\rm crit}) = 0$, which marks the epoch at which the Universe switched from a decelerating to an accelerating phase~\cite{Frieman2008,CaldwellKamionkowski2009,Sahni:2008xx,Farooq:2016zwm}.

\subsubsection*{Physical interpretation}

The acceleration equation reveals that cosmic dynamics is determined by the interplay between the matter density and the properties of dark energy~\cite{Carroll2001,Frieman2008,Copeland2006,CaldwellKamionkowski2009,Amendola:2012ys}. Specifically, acceleration is governed by the combination
\begin{equation}
\rho + 3p = \rho_m + \rho_{\rm DE}\bigl(1+3w_0\bigr),
\end{equation}
showing that acceleration is not a function of $w_0$ or $\Omega_{\rm DE}$ in isolation. When $\Omega_{\rm DE}$ is small, the gravitational pull of matter dominates and even a strongly negative $w_0$ may be insufficient to drive acceleration. When $\Omega_{\rm DE}$ is large, the threshold relaxes toward $w_0 = -1/3$, so acceleration is achievable for a less strongly negative equation of state, though $w_0$ must still lie below this bound.  This non-trivial interplay implies that any inference on the present day acceleration must rely on a simultaneous, well constrained determination of both $\Omega_{\rm DE}$ (or, equivalently, $\Omega_m$) and $w_0$~\cite{Carroll2001,Copeland2006,Caldwell:2005tm,Sahni:2008xx,Clarkson:2007pz}. Within the CPL framework, $w_0$ is the parameter most directly tied to the low redshift expansion history~\cite{Chevallier:2000qy,Linder:2002et,Caldwell:2005tm}, and it is therefore the most sensitive to the inclusion or omission of low-$z$ data, which is the central theme of the comparative analysis presented in the remainder of this paper~\cite{DESI:2025zgx,eBOSS:2020yzd,Sahni:2014ooa,Farooq:2016zwm}.

\begin{table*}[htbp]
\centering
\caption{Median values and 68\% confidence intervals for the cosmological parameters across the four models: $w$CDM, $\nu$CDM ($\Lambda$CDM\,$+\sum m_\nu$), $w$CDM\,$+\sum m_\nu$, and $w_0w_a$CDM (CPL). Dashes (---) denote parameters not present (or fixed) in a given model. For $w_0w_a$CDM, $z_{\rm crit}$ is the transition redshift defined by $q(z_{\rm crit})=0$. The Planck+SDSS($z{>}0.295$) row corresponds to the SDSS sample with the low redshift MGS anchor removed (Sec.~\ref{subsec:noMGS}). We have verified that all quoted $q_0$ values are consistent with equation~(\ref{eq:q0flat}) evaluated at the corresponding median $\Omega_{m,0}$ and $w_0$.}
\label{tab:all_models}
\resizebox{\textwidth}{!}{%
\begin{tabular}{lccccccccc}
\toprule
\textbf{Case} & $\Omega_{m,0}$ & $\sum m_\nu$ [eV] & $w_0$ & $w_a$ & $h_0$ & $q_0$ & $\Omega_{\mathrm{DE},0}$ & Age [Gyr] & $z_{\rm crit}$ \\
\midrule
\multicolumn{10}{c}{\textit{$w$CDM model}} \\
\midrule
Planck                    & $0.19^{+0.06}_{-0.03}$ & ---                    & $-1.59^{+0.34}_{-0.24}$ & ---            & $0.87^{+0.09}_{-0.12}$ & $-1.44^{+0.54}_{-0.39}$ & $0.81^{+0.03}_{-0.06}$ & $13.58^{+0.14}_{-0.07}$ & --- \\
Planck+DESI               & $0.31^{+0.01}_{-0.00}$ & ---                    & $-0.97 \pm 0.02$        & ---            & $0.67 \pm 0.01$         & $-0.51 \pm 0.03$        & $0.69^{+0.00}_{-0.01}$ & $13.83 \pm 0.02$        & --- \\
Planck+SDSS              & $0.30 \pm 0.01$        & ---                    & $-1.04 \pm 0.04$        & ---            & $0.69 \pm 0.01$         & $-0.59 \pm 0.05$        & $0.70 \pm 0.01$         & $13.80^{+0.03}_{-0.02}$ & --- \\
Planck+DESI+PanthPlus     & $0.30 \pm 0.01$        & ---                    & $-1.00 \pm 0.02$        & ---            & $0.68 \pm 0.01$         & $-0.54 \pm 0.03$        & $0.70 \pm 0.01$         & $13.82 \pm 0.02$        & --- \\
\midrule
\multicolumn{10}{c}{\textit{$\nu$CDM model ($\Lambda$CDM\,$+\sum m_\nu$)}} \\
\midrule
Planck                    & $0.32^{+0.02}_{-0.01}$ & $0.07^{+0.09}_{-0.05}$ & ---                     & ---            & $0.67 \pm 0.01$         & $-0.52 \pm 0.02$        & $0.68^{+0.01}_{-0.02}$ & $13.81^{+0.04}_{-0.06}$   & --- \\
Planck+DESI               & $0.30 \pm 0.00$        & $0.02^{+0.02}_{-0.01}$ & ---                     & ---            & $0.68 \pm 0.00$         & $-0.55 \pm 0.01$        & $0.70 \pm 0.00$         & $13.78^{+0.02}_{-0.02}$   & --- \\
Planck+SDSS              & $0.31 \pm 0.01$        & $0.03^{+0.04}_{-0.02}$ & ---                     & ---            & $0.68 \pm 0.00$         & $-0.54 \pm 0.01$        & $0.69 \pm 0.01$         & $13.78^{+0.02}_{-0.02}$      & --- \\
Planck+SDSS+PanthPlus    & $0.31 \pm 0.01$        & $0.03^{+0.04}_{-0.02}$ & ---                     & ---            & $0.68 \pm 0.00$         & $-0.54 \pm 0.01$        & $0.69 \pm 0.01$         & $13.78^{+0.02}_{-0.02}$   & --- \\
\midrule
\multicolumn{10}{c}{\textit{$w$CDM\,$+\sum m_\nu$ model}} \\
\midrule
Planck                    & $0.16 \pm 0.05$        & $0.18 \pm 0.14$ &  $-1.96 \pm 0.38$        & ---            & $0.99 \pm 0.14$         & $-2.00 \pm 0.60$        & $0.84 \pm 0.05$         & $13.51 \pm 0.13$        & --- \\
Planck+DESI               & $0.29 \pm 0.01$        & $0.03^{+0.03}_{-0.02}$ & $-1.04 \pm 0.04$        & ---            & $0.69 \pm 0.01$         & $-0.60 \pm 0.05$        & $0.71 \pm 0.01$         & $13.77 \pm 0.02$        & --- \\
Planck+SDSS              & $0.30 \pm 0.01$        & $0.05^{+0.06}_{-0.03}$ & $-1.03^{+0.04}_{-0.05}$ & ---            & $0.69 \pm 0.01$         & $-0.58^{+0.05}_{-0.06}$ & $0.70 \pm 0.01$         & $13.77 \pm 0.03$        & --- \\
Planck+DESI+PanthPlus     & $0.30 \pm 0.01$        & $0.02^{+0.03}_{-0.01}$ & $-0.99 \pm 0.02$        & ---            & $0.68 \pm 0.01$         & $-0.53 \pm 0.03$        & $0.70 \pm 0.01$         & $13.79 \pm 0.02$        & --- \\
\midrule
\multicolumn{10}{c}{\textit{$w_0w_a$CDM model (CPL)}} \\
\midrule
Planck                    & $0.21^{+0.11}_{-0.06}$ & ---                    & $-1.12^{+0.61}_{-0.53}$ & $-1.71^{+0.99}_{-0.87}$ & $0.83^{+0.17}_{-0.16}$ & $-0.83^{+0.81}_{-0.78}$ & $0.80^{+0.06}_{-0.11}$ & $13.53^{+0.19}_{-0.12}$ & $0.84$ \\
Planck+DESI               & $0.35 \pm 0.02$        & ---                    & $-0.41^{+0.21}_{-0.22}$ & $-1.78^{+0.61}_{-0.61}$ & $0.63 \pm 0.02$         & $0.10^{+0.21}_{-0.23}$  & $0.65 \pm 0.02$         & $13.79 \pm 0.02$        & $0.08,\ 0.86$ \\
Planck+SDSS              & $0.33 \pm 0.02$        & ---                    & $-0.71^{+0.19}_{-0.18}$ & $-0.94^{+0.53}_{-0.59}$ & $0.66 \pm 0.02$         & $-0.22^{+0.20}_{-0.21}$ & $0.67 \pm 0.02$         & $13.77 \pm 0.03$        & $0.76$ \\
% TO VERIFY before submission: row values taken from the authors' no-MGS chain notes / Eq. (19); confirm z_crit against the chain. this is the value z_crit = 0.5245 (-0.1478/+0.1380)
Planck+SDSS($z{>}0.295$) & $0.34 \pm 0.03$        & ---                    & $-0.61^{+0.32}_{-0.31}$ & $-1.18^{+0.81}_{-0.83}$ & $0.65 \pm 0.03$         & $-0.10^{+0.33}_{-0.35}$ & $0.66 \pm 0.03$         & $13.77 \pm 0.03$        & $0.53$ \\
Planck+DESI+PanthPlus     & $0.31 \pm 0.01$        & ---                    & $-0.84^{+0.06}_{-0.05}$ & $-0.61^{+0.20}_{-0.21}$ & $0.67 \pm 0.01$         & $-0.37 \pm 0.06$        & $0.69 \pm 0.01$         & $13.78 \pm 0.02$        & $0.74$ \\
\bottomrule
\end{tabular}%
}
\end{table*}

\section{Methodology and Observational Datasets}\label{sec:Data}

The cosmological constraints and reconstructions presented in this work focus primarily on the Chevallier--Polarski--Linder (CPL) dynamical dark energy framework, parametrized by the pair $(w_0, w_a)$~\cite{Chevallier:2000qy,Linder:2002et}, which we adopt as our baseline model, while the additional models introduced above are considered for comparison. The expansion rate $H(z)$, the deceleration parameter $q(z)$, and the equation of state $w(z)$ are derived self-consistently from this underlying parametrization, ensuring that all reported quantities share a common theoretical basis~\cite{Copeland2006,Frieman2008,DESI:2025zgx}. In what follows, we describe the complementary data combinations employed in our analysis.

\subsection{Planck 2018 Baseline Analysis}

To establish an internally consistent reference, we carried out an independent Markov Chain Monte Carlo (MCMC) analysis of the \textit{Planck} 2018 cosmic microwave background (CMB) measurements~\cite{Planck:2018vyg,Planck:2019nip}. The sampling is performed with the \texttt{Cobaya} framework~\cite{Torrado:2020dgo} interfaced with the \texttt{CLASS} Boltzmann solver~\cite{Lesgourgues:2011re,Blas:2011rf}, in which dark energy is implemented as a fluid component following the CPL parametrization~\cite{Chevallier:2000qy,Linder:2002et}. The baseline likelihood combination includes the low-multipole temperature and polarization spectra (\texttt{TT} and \texttt{EE}), the high-multipole \texttt{Plik} \texttt{TTTEEE} lite likelihood, and the \textit{Planck} CMB lensing reconstruction~\cite{Planck:2018vyg,Planck:2019nip,Planck:2018lensing}. The corresponding likelihood implementations adopted in \texttt{Cobaya} are \texttt{planck\_2018\_lowl.TT}, \texttt{planck\_2018\_lowl.EE}, \texttt{planck\_2018\_highl\_plik.TTTEEE\_lite\_native}, and \texttt{planck\_2018\_lensing.native}. Convergence of the chains is monitored through the Gelman--Rubin diagnostic~\cite{Gelman:1992zz}, with $R-1 < 0.01$ adopted as the convergence criterion~\cite{Lewis:2002ah,Lewis:2019xzd}. These chains provide the \textit{Planck}-only rows of Table~\ref{tab:all_models}; the combined rows instead use the externally released chains described below, which adopt different CMB likelihood and lensing choices (see the caveat at the end of this section).

\subsection{Planck + SDSS DR16 Combined Analysis} 

For the joint \textit{Planck}+SDSS analysis, rather than re-running the full likelihood pipeline, we make use of the official DR16 cosmological MCMC chains publicly released by the SDSS collaboration~\cite{eBOSS:2020yzd}.\footnote{Specifically, we employ the \texttt{base\_w\_wa\_CMBLens\_BAORSD} chains available at 
\url{https://svn.sdss.org/public/data/SDSS/DR16cosmo/tags/v1_0_1/mcmc/base_w_wa/CMBLens_BAORSD/}.} These chains combine the \textit{Planck} CMB lensing reconstruction~\cite{Planck:2018vyg,Planck:2018lensing} with Baryon Acoustic Oscillation (BAO) and Redshift Space Distortion (RSD) measurements from SDSS DR16~\cite{eBOSS:2020yzd,Bautista:2020ahg,duMasDesBourboux:2020pck}, considering a broad redshift range and providing tight constraints on the late-time expansion history. The resulting dataset is used to constrain a time-varying equation of state of the form $w(a) = w_0 + w_a(1-a)$~\cite{Chevallier:2000qy,Linder:2002et}, as it breaks the geometric degeneracies that affect CMB-only analyses through the inclusion of low redshift distance and growth information.

\subsection{Planck + SDSS DR16 + Pantheon+}
This combination is taken from the same SDSS repository and, as released there, is available for the $\nu$CDM case~\cite{eBOSS:2020yzd,Planck:2018vyg,Planck:2018lensing,Lesgourgues:2006nd,Lattanzi:2017ubd}. We therefore use it only as a $\nu$CDM comparison; a CPL Planck+SDSS+Pantheon+ chain is not part of the public release and is not constructed here, so the corresponding CPL entry is omitted from Table~\ref{tab:all_models} (cf.\ the symmetric Planck+DESI+Pantheon+ row, which is publicly available in CPL form).

\subsection{DESI Constraints}

For the analyses based on the Dark Energy Spectroscopic Instrument (DESI), we make use of the publicly available Year-3 BAO cosmology chains released by the DESI collaboration~\cite{DESI:2025zgx}.\footnote{Available at \url{https://data.desi.lbl.gov/public/papers/y3/bao-cosmo-params/cobaya/base_w_wa/}.} Among the dataset configurations distributed in the public repository, we focus on two combinations that are best matched to the scope of the present work:

\begin{itemize}
    \item \textit{DESI+CMB}: combines the DESI DR2 (Year-3) BAO measurements~\cite{DESI:2025zgx,DESI:2025fii} with the \textit{Planck} 2018 primary CMB likelihoods~\cite{Planck:2018vyg,Planck:2019nip} (low-$\ell$ \texttt{TT} and \texttt{EE}, together with the NPIPE \texttt{CamSpec} high-$\ell$ \texttt{TTTEEE} spectra~\cite{Rosenberg:2022sdy}) and the joint \textit{Planck}+ACT~DR6 CMB lensing reconstruction~\cite{Madhavacheril:2023dr6}.\footnote{Chain identifier: \texttt{desi-bao-all\_planck2018-lowl-TT-\\ clik\_planck2018-lowl-EE-clik\_planck-NPIPE-highl-Cam\\Spec-TTTEEE\_planck-act-dr6-lensing}.}
    
    \item \textit{DESI+CMB+PantheonPlus}: extends the previous combination by including the Pantheon+ compilation of Type~Ia supernovae~\cite{Scolnic:2021amr,Brout:2022vxf}, which adds an independent low redshift distance probe and substantially tightens the constraints on the late-time expansion history.\footnote{Chain identifier: \texttt{desi-bao-all\_pantheonplus\_planck\\2018-lowl-TT-clik\_planck2018-lowl-EE-clik\_planck-\\NPIPE-highl-CamSpec-TTTEEE\_planck-act-dr6-lensing}.}
\end{itemize}

All DESI chains were produced with the \texttt{Cobaya} sampler~\cite{Torrado:2020dgo} within the same CPL $(w_0, w_a)$ framework adopted throughout this work~\cite{Chevallier:2000qy,Linder:2002et,DESI:2025zgx}, so that the comparison between the SDSS DR16 and DESI DR2 constraints is carried out within a common dark energy parametrization and is not affected by parametrization-induced biases~\cite{eBOSS:2020yzd,DESI:2025zgx,Wolf:2025robustness,Lee:2025shape}.

\paragraph{Caveat on the CMB and lensing inputs.} We stress that the ``\textit{Planck}'' information is \emph{not} identical across the three combinations used here, and our comparison should be read with this in mind. The \textit{Planck}-only rows come from our runs made with the 2018 \texttt{Plik} \texttt{TTTEEE} lite high-$\ell$ likelihood and the 2018 lensing reconstruction; the DESI combinations use the NPIPE \texttt{CamSpec} high-$\ell$ likelihood together with the joint \textit{Planck}+ACT~DR6 lensing; and the SDSS DR16 combination, as publicly released, pairs the \textit{Planck} CMB information with BAO and RSD but with the 2018-era likelihood and lensing rather than the NPIPE/ACT~DR6 inputs. The SDSS chains additionally include RSD (growth) information, which the DESI BAO chains do not. These differences in the CMB likelihood version, in the lensing dataset, and in the inclusion of RSD are subdominant to the difference in BAO redshift coverage that is the focus of this work, but they are not zero: part of any residual offset between the SDSS and DESI columns may originate here rather than from the BAO data themselves. The same analysis with a single, matched CMB+lensing likelihood would isolate the BAO-only contribution and is the natural next step. A cheaper intermediate check would repeat the comparison with the corresponding BAO-only (no-RSD) SDSS chains, also part of the DR16 public release, thereby isolating the purely geometric information. We attempt neither here, and we therefore frame our conclusions in terms of the \emph{direction} and approximate size of the shifts rather than their precise statistical significance.

\begin{figure*}[htb]
\centering
\begin{minipage}{1.0\columnwidth}
    \centering
    \includegraphics[width=\linewidth]{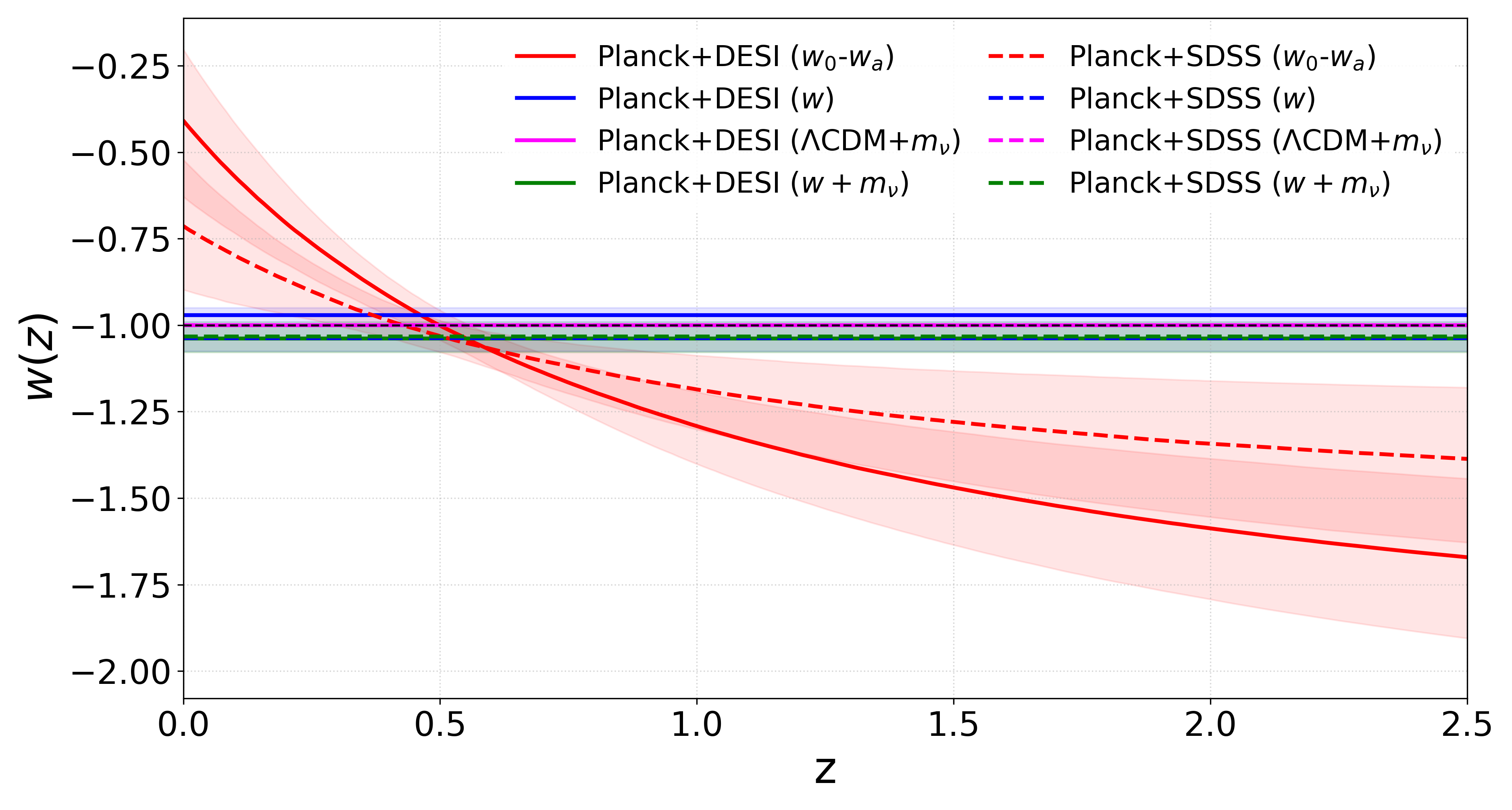}
\end{minipage}
\hfill
\begin{minipage}{1.0\columnwidth}
    \centering
    \includegraphics[width=\linewidth]{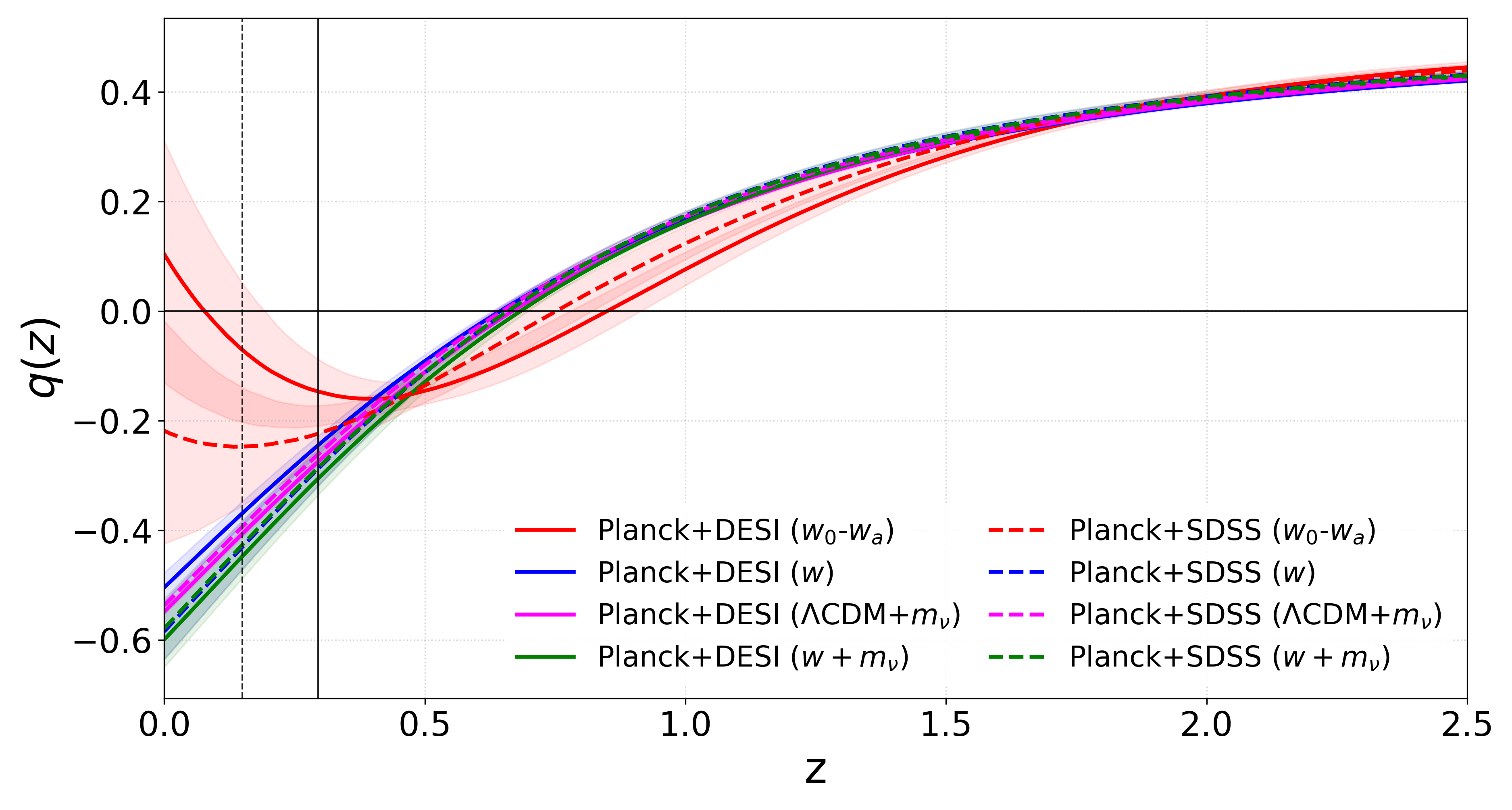}
\end{minipage}

\caption{
Comparison of the equation of state $w(z)$ (left) and the deceleration parameter $q(z)$ (right) across the different models considered for the two main BAO combinations. Four models are shown: $w_0w_a$CDM (red), $w$CDM (blue), $\nu$CDM ($\Lambda$CDM\,$+\sum m_\nu$) (pink), and $w$CDM\,$+\sum m_\nu$ (green). Planck+DESI is shown with continuous lines and Planck+SDSS with dashed lines. The black vertical lines indicate the smallest effective redshift probed by each dataset (continuous for Planck+DESI, dashed for Planck+SDSS).
} 
\label{fig:fig1}
\end{figure*}

\begin{figure*}
\centering
\begin{minipage}{1.2\columnwidth}
    \centering
    \includegraphics[width=\linewidth]{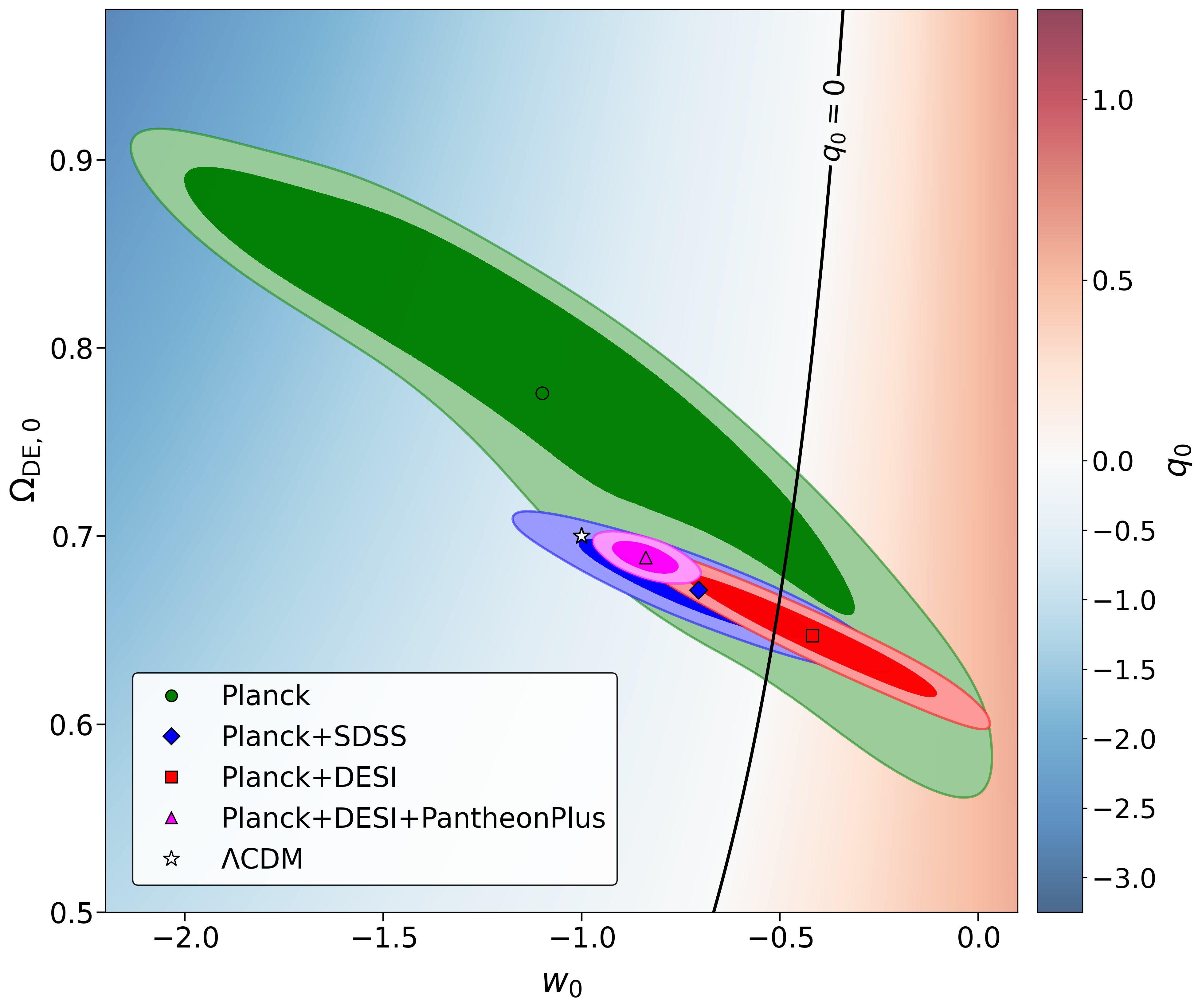}
\end{minipage}
\caption{Two-dimensional posterior distribution in the $(w_0,\Omega_{\rm DE,0})$ plane, color-coded by the present day deceleration parameter $q_0$. The solid black curve corresponds to $q_0 = 0$ and divides the plane into an accelerating region ($q_0 < 0$, at more negative $w_0$ and larger $\Omega_{\rm DE,0}$, to the left of the curve) and a decelerating region ($q_0 > 0$, to the right of the curve), highlighting the interplay between the dark energy equation of state $w_0$ and its present day fractional density $\Omega_{\rm DE,0}$ in driving cosmic acceleration.}
\label{fig:fig2}
\end{figure*}

%%%%%%%%%%%%%%%%%%%%%%%%%%%%

%%%%%%%%%%%%%%%%%%%%%%%%%%%%%%%%%%

%%%%%%%%%%%%%%%%%%%%%%%%%%%%%%%%%%%%%%%%%

%%%%%%%%%%%%%%%%%%%%%%%%%%%%%%%%%%%%%%%%%%%%%%%
\section{Results}\label{sec:Results}

In this section we present the constraints obtained for the four cosmological models considered in this work, $w$CDM, $\nu$CDM, $w$CDM $+\sum m_\nu$, and the 
Chevallier--Polarski--Linder $w_0w_a$CDM parametrization, using the dataset combinations \textit{Planck} alone, \textit{Planck}+SDSS, \textit{Planck}+DESI, and \textit{Planck}+DESI+Pantheon+. 
The analysis is organized around three diagnostics of late-time dynamics: the equation of state 
$w(z)$, the deceleration parameter $q(z)$, and the joint $q_0$--$w_0$ relation~\cite{Carroll2001,Frieman2008,Copeland2006,CaldwellKamionkowski2009}.  All numerical values quoted below are collected in Table~\ref{tab:all_models}, and the corresponding reconstructions are shown in Fig.~\ref{fig:fig1}.

\subsection{Equation of State Evolution}

Figure~\ref{fig:fig1} (left panel) shows the reconstructed dark energy 
equation of state $w(z)$ for the cosmological models considered in this work and for the different dataset combinations~\cite{Carroll2001,Frieman2008,Copeland2006}.
As expected, the $w$CDM, $\nu$CDM and $w$CDM\,$+\sum m_\nu$ models display a constant (or 
effectively constant) equation of state across the entire redshift range: 
$\nu$CDM has $w = -1$ by construction, while the two $w$CDM 
variants return values clustered tightly around $w \simeq -1$ for every 
dataset combination. Only the 
$w_0w_a$CDM (CPL) parametrization produces a genuinely evolving $w(z)$, 
and it is therefore in this model that the differences between BAO 
samples become physically meaningful.

% TO VERIFY before submission: confirm that z_pd and z_crit are indeed evaluated at the
% posterior medians (the SDSS z_pd = 0.437 differs slightly from the value obtained by
% plugging in the median (w0, wa) = (-0.71, -0.94), which gives 0.446).       ####ACF if you plug the median with more significative values it comes! 

Within CPL, the \textit{Planck}+SDSS combination yields a quintessence-like behavior at low redshift, with the reconstructed $w(z)$ crossing the 
phantom divide $w=-1$ at $z_{\rm pd}^{\rm SDSS} \simeq 0.437$ (here and in the following, derived quantities such as $z_{\rm pd}$ and $z_{\rm crit}$ are evaluated at the posterior median parameters) and reaching 
a present day value
\begin{equation}
    w_0^{\rm SDSS} = -0.71^{+0.19}_{-0.18}\, .
\end{equation}
The \textit{Planck}+DESI combination, by contrast, crosses $w=-1$ at a 
substantially higher redshift, $z_{\rm pd}^{\rm DESI} \simeq 0.498$, and 
evolves much more steeply toward less negative values, returning at 
$z=0$
\begin{equation}
    w_0^{\rm DESI} = -0.41^{+0.21}_{-0.22}\, .
\end{equation}

The DESI inferred present day equation of state is therefore $\sim\!0.3$ \emph{higher} (i.e.\ less negative, further from the cosmological constant 
value) than the SDSS one. Treating the two determinations as independent, this offset corresponds to only $\approx 1.1\sigma$ (a difference of $0.30$ against a combined uncertainty $\sqrt{0.21^2+0.19^2}\simeq0.28$); it is thus a marginal, not a decisive, difference, and we interpret it as a directional trend rather than a significant tension; the corresponding offset in $q_0$ ($0.32$ against $\sqrt{0.22^2+0.21^2}\simeq0.30$) is similarly $\approx1.1\sigma$. We note, moreover, that because the two combinations share the \textit{Planck} likelihood, the common information partially cancels in the difference and the two posteriors are positively correlated; the independence assumption therefore overestimates the variance of the difference, so $\approx1.1\sigma$ should be read as a conservative estimate of the BAO-driven offset. The shift reflects the freedom of the CPL parameters to drift in the absence of low redshift 
anchors below $z\!\sim\!0.295$. A rigorous tension quantification accounting for the shared CMB information and parameter correlations would require the joint posterior of the two analyses, which is beyond the scope of the present, chain-level comparison.

The addition of 
Pantheon+ to \textit{Planck}+DESI pulls the present day value down to 
$w_0 = -0.84^{+0.06}_{-0.05}$ and pushes the crossing closer to 
$z_{\rm pd} \simeq 0.358$. The inclusion of low-$z$ 
information drives $w_0$ toward more negative values, in the direction of 
$\Lambda$CDM~\cite{Scolnic:2021amr,Brout:2022vxf}.

This difference in $w_0$ has a direct impact on the deceleration parameter 
through the relation $q_0 = \tfrac{1}{2}\Omega_{m,0}+ \tfrac{1}{2}(1+3w_0)\,\Omega_{\mathrm{DE},0}$ ~\cite{Carroll2001,Frieman2008,Copeland2006}: a less negative $w_0$, as preferred by Planck+DESI alone, mechanically pushes $q_0$ toward positive values 
and can place the median posterior on the non-accelerating side of the $q_0=0$ boundary.  We quantify this effect in the next subsection.

\subsection{Deceleration Parameter}

The right panel of Fig.~\ref{fig:fig1} shows the reconstructed deceleration parameter $q(z)$~\cite{Carroll2001,Frieman2008,Copeland2006,CaldwellKamionkowski2009}.
Cosmic acceleration occurs when $q(z) < 0$; the key results are collected in Table~\ref{tab:all_models}.

\begin{figure*}[htb]
\centering
\begin{minipage}{0.49\textwidth}
    \centering
    \includegraphics[width=\linewidth]{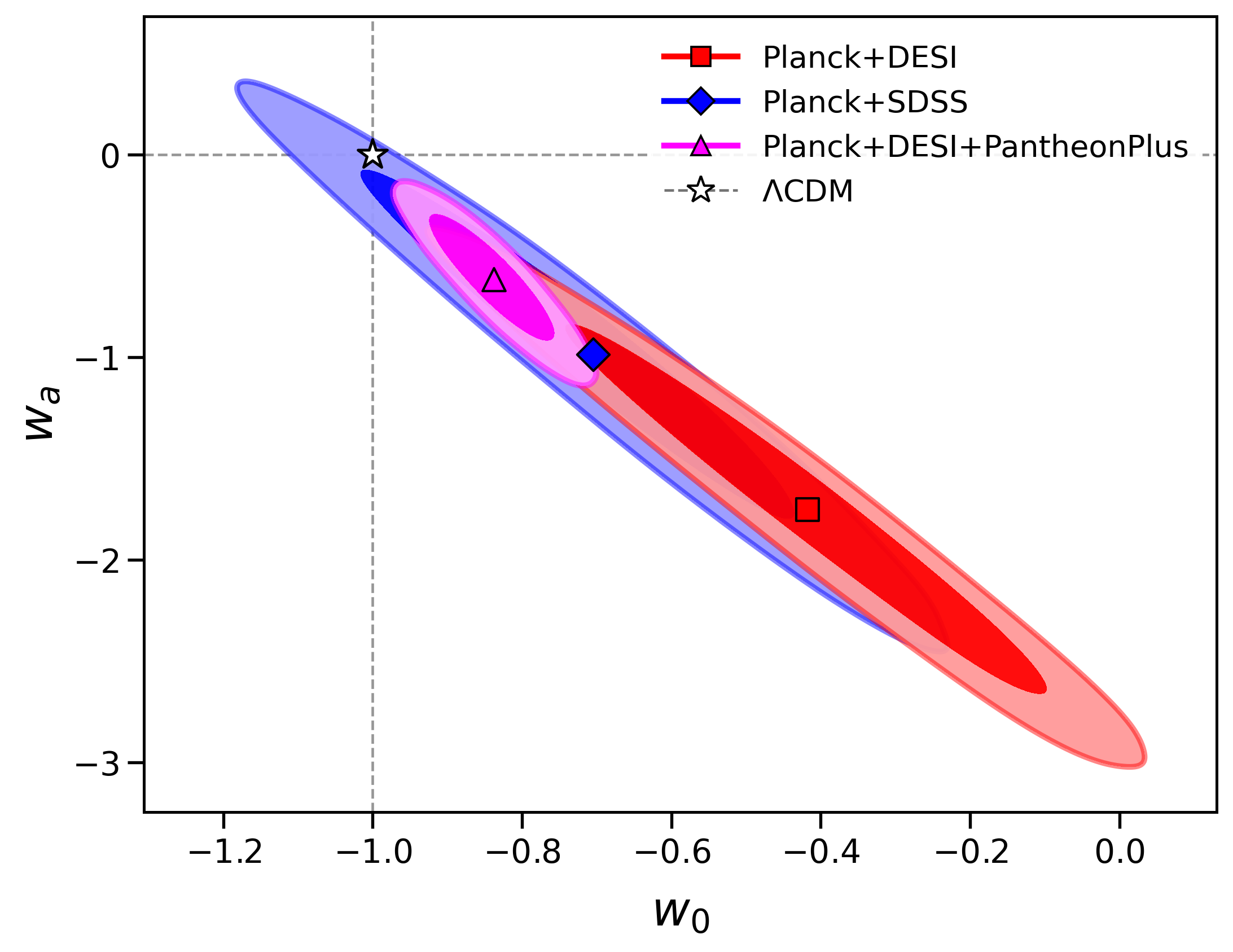}
\end{minipage}
\hfill
\begin{minipage}{0.49\textwidth}
    \centering
    \includegraphics[width=\linewidth]{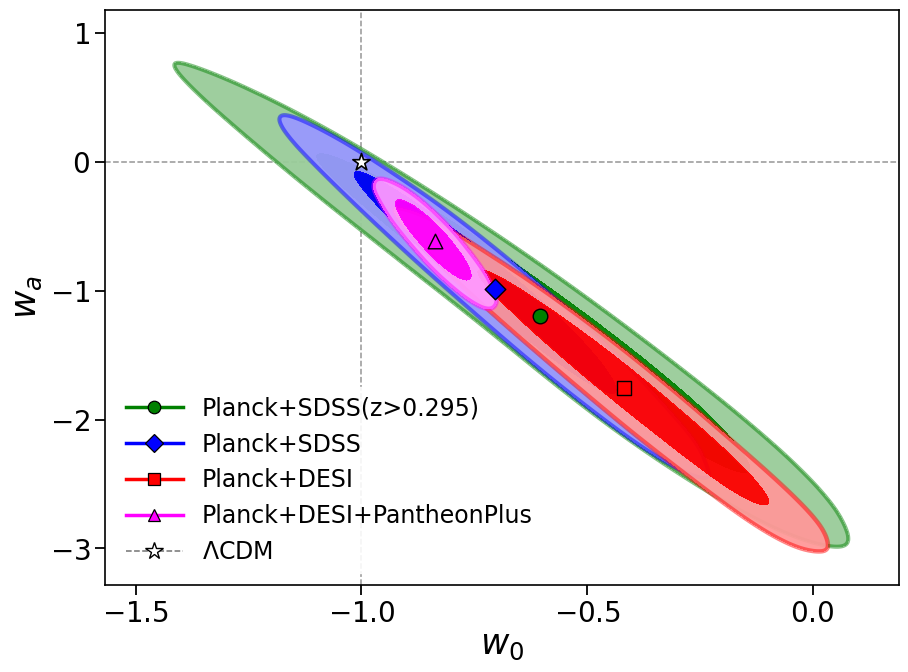}
\end{minipage}
\caption{Marginalized posteriors in the $(w_0,w_a)$ plane for three data
combinations. \textbf{Left:} using the full SDSS DR16 sample. \textbf{Right:}
with the low redshift MGS anchor removed. Switching the BAO dataset from SDSS to
DESI shifts the contours away from the cosmological-constant point
$(w_0,w_a)=(-1,0)$, while the further inclusion of Pantheon+ decreases tension 
with $\Lambda$CDM. Removing the MGS anchor (right) broadens the SDSS posterior
and drifts the best fit toward the DESI contour, illustrating the role of low redshift
coverage.}
\label{fig:fig3}
\end{figure*}

\begin{figure*}[htb]
\centering
\begin{minipage}{1.0\columnwidth}
    \centering
    \includegraphics[width=\linewidth]{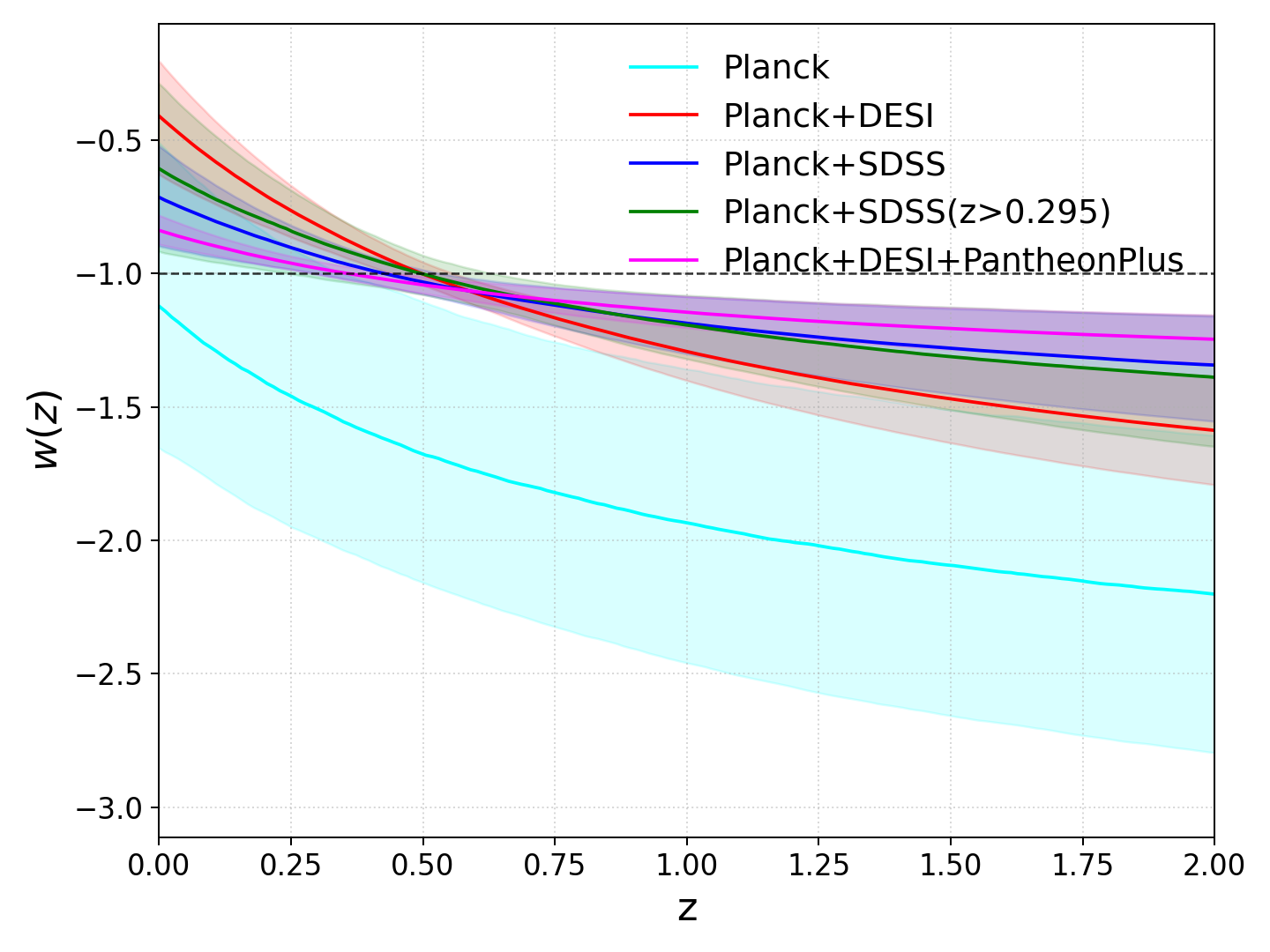}
\end{minipage}
\hfill
\begin{minipage}{1.0\columnwidth}
    \centering
    \includegraphics[width=\linewidth]{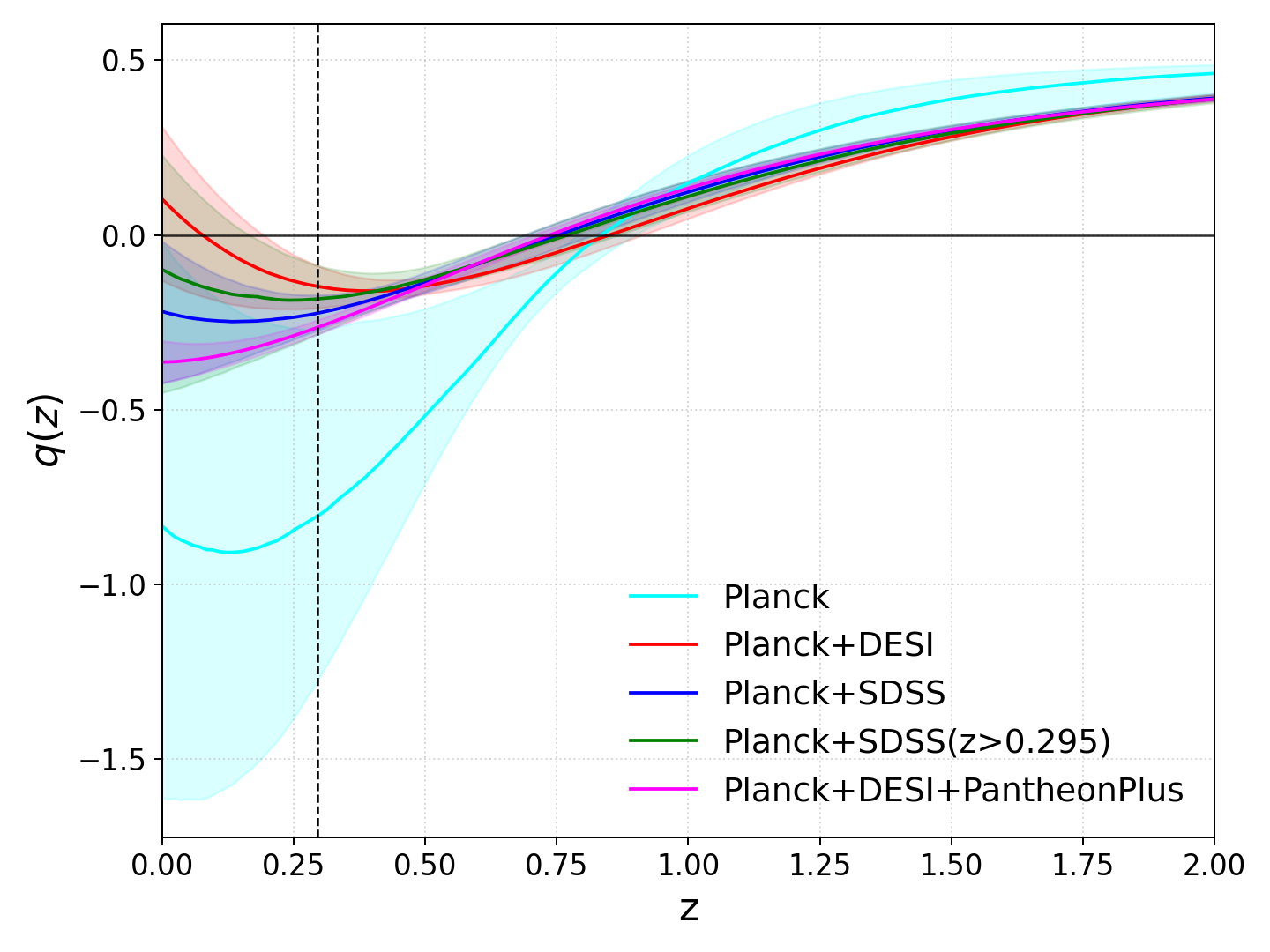}
\end{minipage}
\caption{
Comparison of the equation of state $w(z)$ (left) and the deceleration parameter $q(z)$ (right) across all dataset combinations, for the model $w_0w_a$CDM. Planck in cyan, Planck+DESI in red, Planck+SDSS in blue, Planck+SDSS($z>0.295$) in green, Planck+DESI+Pantheon+ in magenta. The black dashed vertical line indicates the smallest effective redshift probed by the Planck+SDSS($z>0.295$) dataset, in which the low redshift SDSS Main Galaxy Sample (MGS) has been removed, in contrast to the dashed line in Fig.~\ref{fig:fig1}.
}
\label{fig:fig4}
\end{figure*}
These results can be summarized as follows:

\begin{itemize}
\item \textit{Planck} alone predicts acceleration ($q_0 < 0$) with median value $q_0 = -0.83^{+0.81}_{-0.78}$, but with large uncertainty.
\item \textit{Planck}+DESI gives a median value of $q_0 = 0.10^{+0.21}_{-0.23}$, indicating that the data cannot distinguish between a presently accelerating and a non-accelerating Universe.
\item \textit{Planck}+SDSS prefers $q_0 < 0$ with median value $q_0 = -0.22^{+0.20}_{-0.21}$ and a well defined transition redshift,
      providing clear evidence for late-time acceleration.
\item \textit{Planck}+DESI+Pantheon+ restores $q_0 < 0$ with median value $q_0 = -0.37^{+0.06}_{-0.06}$, confirming that the inclusion
      of supernova data is crucial for establishing acceleration when using DESI.
\end{itemize}

As shown in Fig.~\ref{fig:fig2}, the \textit{Planck}-only analysis yields very
broad contours in the $w_0$--$\Omega_{\rm DE,0}$ plane, reflecting the well known
geometric degeneracy: since the CMB constrains dark energy only through the
angular diameter distance to last scattering, a wide range of $w_0$ and
$\Omega_{\rm DE,0}$ combinations reproduce the same distance and are therefore
indistinguishable to \textit{Planck} alone~\cite{Planck:2018vyg}. The inclusion of BAO information substantially tightens the posterior, although the qualitative behavior depends on which BAO dataset is combined with \textit{Planck}. When SDSS DR16 BAO+RSD measurements are added, the median value of the present day deceleration parameter is $q_0 = -0.22$, and the bulk of the posterior remains within the accelerating regime ($q_0 < 0$)~\cite{eBOSS:2020yzd,Carroll2001,Frieman2008,Copeland2006}. In contrast, replacing SDSS with the DESI BAO measurements shifts the central value to $q_0 = +0.10$, placing the median and most of the allowed region within the decelerating regime ($q_0 > 0$) a difference between the two BAO datasets that, as quantified above, amounts to only $\approx1\sigma$ when interpreted in the CPL framework. The further inclusion of the Pantheon+ Type~Ia supernova compilation in the \textit{Planck}+DESI combination drives $q_0$ back to $-0.37$, restoring an accelerating present day expansion that is in good agreement with the value inferred from \textit{Planck}+SDSS. This sequence highlights the strong sensitivity of $q_0$ to the specific late-time distance probes employed, and the key role played by SNe Ia in anchoring the low redshift expansion history~\cite{Scolnic:2021amr,Brout:2022vxf,DESI:2025zgx}.

\begin{figure*}
\centering
\begin{minipage}{\textwidth}
    \centering
    \includegraphics[width=\linewidth]{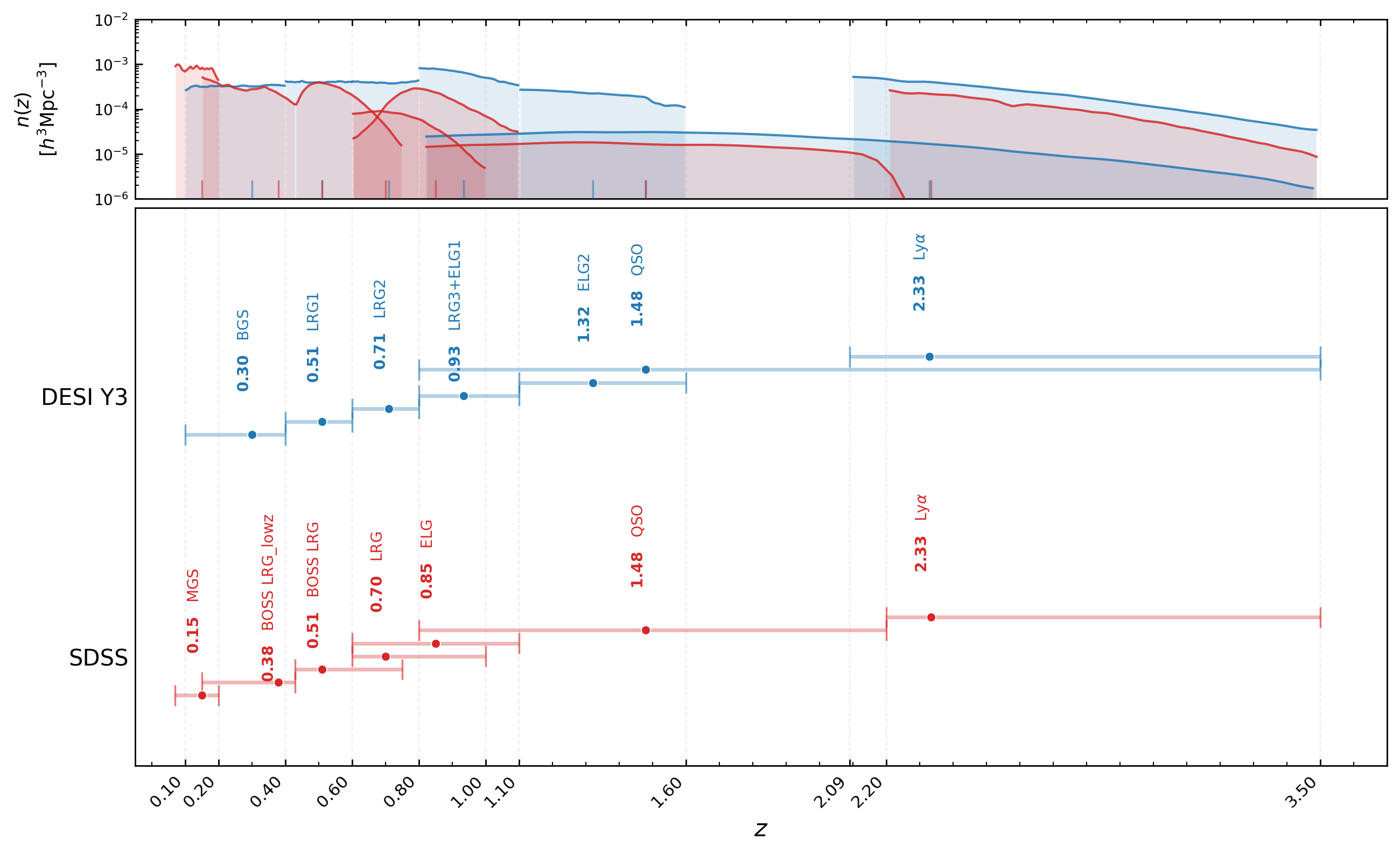}
\end{minipage}
\caption{Visualization of the BAO tracers in the redshift domain. Each point marks the effective redshift $z_{\rm eff}$ of a tracer bin: blue for the DESI DR2 (Year-3) sample and red for the SDSS sample. The upper panel shows the redshift distribution of the number density $n(z)$ of the tracers per unit volume in each bin. The lowest redshift DESI point (the BGS bin at $z_{\rm eff}\approx0.295$) sits well above the SDSS MGS point at $z_{\rm eff}\approx0.15$, illustrating the gap in low redshift coverage discussed in the text.}
\label{fig:fig5}
\end{figure*}

\section{Discussion}\label{sec:Discussion}
\subsection{Scope: present day acceleration versus dynamical dark energy}

Before comparing the models, we clarify the scope of our claim, since two logically distinct questions are easily confused. The first is whether the Universe is \emph{currently accelerating}, i.e.\ the sign of $q_0$; this is a present day quantity fixed almost entirely by $w_0$ and $\Omega_{\rm DE,0}$ and is therefore acutely sensitive to the lowest redshift at which the expansion is anchored. The second is whether dark energy is \emph{dynamical}, i.e.\ whether $w_a\neq0$; this is the property the DESI collaboration reports at high significance, and crucially, that preference is driven largely by the intermediate-redshift LRG bins and survives the inclusion of supernovae. Our analysis speaks primarily to the \emph{first} question. We argue that the BAO+CMB-only inference of $q_0$ (and of $w_0$) from DESI is fragile because it lacks a $z_{\rm eff}<0.3$ anchor, and that this fragility, rather than new physics, can account for the apparent preference for a non-accelerating present epoch in that particular combination. It does \emph{not}, by itself, explain away the full $w_0w_a$CDM preference obtained when low redshift supernovae are included; that result already carries low redshift information and must be addressed on its own terms. We return to this distinction in Sec.~\ref{sec:Conclusions}.

\subsection{Model Constraints and the $w_0w_a$CDM Case}

The results summarized in Table~\ref{tab:all_models} allow us to compare the four cosmological models analyzed in this work, $w$CDM, $\nu$CDM, $w$CDM$+\sum m_\nu$, and the Chevallier--Polarski--Linder parametrization $w_0w_a$CDM~\cite{Carroll2001,Chevallier:2000qy,Linder:2002et}, in terms of the different datasets~\cite{Planck:2018vyg,eBOSS:2020yzd,DESI:2025zgx,Scolnic:2021amr,Brout:2022vxf}. 
For the single-parameter extensions of $\Lambda$CDM, $w$CDM and $\nu$CDM, the recovered values of $\Omega_{m,0}$, $h_0$, and $q_0$ are consistent at the $1\sigma$ level with BAO information included, and the late-time dynamical picture is essentially unchanged whether or not $\sum m_\nu$ is varied~\cite{Lesgourgues:2006nd,Hannestad:2005gj,Planck:2018vyg}. In other words, neither a constant deviation of $w$ from $-1$ nor the addition of a free neutrino mass produces a qualitative departure from the $\Lambda$CDM expectation: the present day expansion remains accelerating, with $q_0$ clustered around its $\Lambda$CDM value, and the different dataset combinations yield posteriors that largely overlap (refer to Fig.~\ref{fig:fig1})~\cite{Carroll2001,Frieman2008,Copeland2006,Zhao:2017cud,Wang:2018fyk}.
This stability is broken as soon as the dark energy sector is allowed a time-dependent equation of state. In the $w_0w_a$CDM case, the additional freedom along the $w_a$ direction couples non-trivially to the late-time distance and growth observables; as a result, the constraints become noticeably more sensitive to the specific low redshift dataset adopted~\cite{Chevallier:2000qy,Linder:2002et,DESI:2025zgx}. The \textit{Planck}+SDSS and \textit{Planck}+DESI combinations return very different inferences for $w_0$ and $q_0$, with the two posteriors located around the boundary between the accelerating and decelerating regimes. Because this is the only model in our analysis where such a qualitative discrepancy emerges, and because it directly bears on the recent debate surrounding the DESI Year-3 hints of dynamical dark energy~\cite{DESI:2025zgx,Cortes:2024lgw,Giare:2025overview}, we devote the remainder of this section to a detailed discussion of the $w_0w_a$CDM results.

\subsubsection{Is the Universe currently accelerating in CPL+Planck+DESI?}
A central finding of this work is that, within the $w_0w_a$CDM 
parametrization~\cite{Chevallier:2000qy,Linder:2002et,Wolf:2025robustness}, the \textit{Planck}+DESI combination \emph{cannot 
unambiguously establish that the Universe is presently accelerating}.  The inferred values

\begin{equation}
    w_0 = -0.41^{+0.21}_{-0.22},~
    q_0 = +0.10^{+0.21}_{-0.23},~\Omega_{\mathrm{DE},0} = 0.65 \pm 0.02,
\end{equation}

yield a positive deceleration parameter at the median, with a $1\sigma$ posterior that straddles the deceleration--acceleration boundary $q_0 = 0$.  In other words, the median DESI prediction is that of a Universe that is \emph{not} currently accelerating, and the reconstruction even returns \emph{two} transition redshifts, $z_{\rm crit} = \{0.08;\,0.86\}$, which suggest a brief acceleration epoch that has already ended in the very recent past. We caution that these transition redshifts are quoted at the posterior median and without uncertainties: the double-root structure arises from a near-tangency of $q(z)$ with zero (see Sec.~\ref{subsec:degeneracy}) and should be read as a qualitative feature of the median reconstruction rather than as a robust detection of a recent end of acceleration.

By contrast, the \textit{Planck}+SDSS combination yields
\begin{equation}
    w_0 = -0.71^{+0.19}_{-0.18},~q_0 = -0.22^{+0.20}_{-0.21},~
    \Omega_{\mathrm{DE},0} = 0.67 \pm 0.02,
\end{equation}

with a single transition redshift $z_{\rm crit} = 0.76$. 
Here, the posterior is centered on a mildly quintessence-like behavior and a genuinely accelerating present epoch.  Although the uncertainties on $q_0$ remain broad, the bulk of the posterior weight lies in the accelerating regime, providing a much clearer signature of late-time acceleration than the Planck+DESI case~\cite{eBOSS:2020yzd,DESI:2025zgx}.

The shift in $w_0$ between the two combinations exceeds $1\sigma$, and the implied values of $q_0$ fall on opposite sides of the acceleration threshold.  
Within the \emph{same} CPL parametrization~\cite{Chevallier:2000qy,Linder:2002et,Wolf:2025robustness,Lee:2025shape}, and with closely similar (though, as discussed in Sec.~\ref{sec:Data}, not identical) CMB information~\cite{Planck:2018vyg,Giare:2024beyond}, the two BAO datasets therefore lead to qualitatively different conclusions about the current dynamical state of the Universe.

\subsubsection{Why DESI is less conclusive: the role of low redshift coverage}

We argue that this result is not a feature of the underlying cosmology but a
direct consequence of the redshift domain spanned by each BAO sample~\cite{Ross:2014qpa,eBOSS:2020yzd,DESI:2024uvr,DESI:2024mwx,DESI:2025zgx,Giare:2024robust,Giare:2025overview,Gu:2025desidr2}. The key
point is the lowest \emph{effective} (mean) redshift at which each survey places
a distance anchor, rather than \emph{the absolute minimum} redshift of any galaxy in the catalogue. These effective anchors differ substantially: the SDSS Main
Galaxy Sample (MGS) provides a measurement at $z_{\rm eff} \approx 0.15$~\cite{Ross:2014qpa,eBOSS:2020yzd},
whereas the lowest-redshift bin of the DESI Bright Galaxy Sample (BGS) has
an effective redshift of only $z_{\rm eff} \approx 0.295$~\cite{DESI:2024uvr,DESI:2024mwx,DESI:2025zgx} (refer to Fig.~\ref{fig:fig5}).

This ${\sim}\,0.15$ difference in effective redshift in the very late Universe
is crucial for $w_0w_a$CDM~\cite{Chevallier:2000qy,Linder:2002et,Giare:2024robust,Wolf:2025robustness,Lee:2025shape}, where two free parameters describe the dark energy
evolution: without a distance anchor below the BGS effective redshift, the
reconstruction must \emph{extrapolate} the equation of state into the epoch
where dark energy dominates the dynamics~\cite{Carroll2001,Frieman2008,Giare:2025overview,Giare:2024beyond,Gu:2025desidr2}.

This extrapolation is precisely what allows the CPL parameters to drift toward
configurations with $w_0$ closer to zero and $\Omega_{m,0}$ as large as $0.35$,
both of which push $q_0$ toward positive values. \textit{Planck}+DESI does not
say that the Universe is decelerating today but it says that, with the data
currently available, one cannot distinguish acceleration from deceleration at
$z = 0$. The \textit{Planck}+SDSS combination, anchored to a lower effective
redshift by the MGS, constrains the very late-time expansion more directly and
recovers a posterior compatible with a presently accelerating Universe~\cite{Ross:2014qpa,eBOSS:2020yzd,Carroll2001,Frieman2008,Giare:2025overview}.

The role of low redshift information is further confirmed by adding the
Pantheon+ supernova compilation to \textit{Planck}+DESI~\cite{Scolnic:2021amr,Brout:2022vxf,Giare:2024robust,Giare:2025overview}. Because the
supernovae extend to much lower effective redshifts, their inclusion bridges the
gap left by the BGS anchor and yields, within the same CPL framework,
\begin{equation}
    w_0 = -0.84^{+0.06}_{-0.05},~
    q_0 = -0.37 \pm 0.06,~\Omega_{\mathrm{DE},0} = 0.69 \pm 0.01,
\end{equation}
together with a single transition redshift $z_{\rm crit} = 0.74$. A clear
accelerating regime is restored, and the uncertainties on $w_0$ and $q_0$ shrink
by nearly a factor of four. This is a direct demonstration that the inconclusive Planck+DESI result reflects missing low redshift leverage rather than any
intrinsic preference of the data for a non-accelerating Universe~\cite{DESI:2025zgx,Scolnic:2021amr}.

\subsubsection{Removing the low redshift MGS anchor from the SDSS sample}\label{subsec:noMGS}

To test this interpretation directly, we repeat the \textit{Planck}+SDSS
analysis with the low redshift SDSS Main Galaxy Sample (MGS) removed~\cite{Ross:2014qpa,eBOSS:2020yzd} (Figs.~\ref{fig:fig3} and~\ref{fig:fig4}, and Table~\ref{tab:all_models}), so that the
remaining sample spans an effective-redshift range more comparable to that of
DESI. Within the same CPL framework we obtain
\begin{equation}
    w_0 = -0.61^{+0.32}_{-0.31},~
    q_0 = -0.10^{+0.33}_{-0.35},~ 
    \Omega_{\mathrm{DE},0} = 0.66 \pm 0.03.
\end{equation}
Removing the low redshift anchor shifts the central value of $q_0$ upward, from
$q_0 = -0.22$ in the full \textit{Planck}+SDSS combination to $q_0 = -0.10$, and
broadens the posteriors considerably, exactly the behavior expected once the
late-time expansion must be reconstructed by extrapolation rather than measured
directly. The result therefore moves towards the \textit{Planck}+DESI value,
without reproducing it exactly: the recovered $q_0$ and $w_0$ lie between the
full-SDSS and DESI cases. This intermediate outcome confirms that the upward
drift of $q_0$ is driven by the loss of low redshift leverage rather than by any
intrinsic difference between the two surveys, while the residual offset reflects
the remaining differences in redshift sampling and survey geometry~\cite{Ross:2014qpa,eBOSS:2020yzd,DESI:2024uvr,DESI:2025zgx}.

\subsubsection{The $\Omega_{\rm DE}$--$w_0$ degeneracy as the underlying driver}\label{subsec:degeneracy}

The contrast between the CPL\,+\,\textit{Planck}+SDSS and
CPL\,+\,\textit{Planck}+DESI combinations is a textbook manifestation of the
$\Omega_{\rm DE}$--$w_0$ degeneracy already discussed in Sec.~\ref{sec:3}~\cite{Chevallier:2000qy,Linder:2002et,DiValentino:2020evt}.
Recalling that the active gravitational source is
$\rho + 3p = \rho_m + \rho_{\mathrm{DE}}(1+3w_0)$, acceleration today requires
\begin{equation}
    w_0 \,<\, -\frac{1}{3\,\Omega_{\mathrm{DE},0}}\,.
\end{equation}
For the \textit{Planck}+DESI combination, $\Omega_{\mathrm{DE},0} = 0.65$ gives
a threshold $w_0 < -0.513$, yet the median $w_0^{\rm Planck+DESI} = -0.41$ lies
\emph{above} this critical value, naturally producing $q_0 > 0$. For the
\textit{Planck}+SDSS combination, $\Omega_{\mathrm{DE},0} = 0.67$ shifts the
threshold to $w_0 < -0.498$, a condition comfortably satisfied by
$w_0^{\rm Planck+SDSS} = -0.71$. The CPL \textit{Planck}+DESI chains therefore
land in the corner of parameter space where dark energy is both too dilute and
too far from the phantom regime to overcome the gravitational pull of matter at
$z = 0$, whereas the CPL \textit{Planck}+SDSS chains sit firmly on the
accelerating side of the same degeneracy direction~\cite{CaldwellKamionkowski2009,DiValentino:2020evt,Carloni:2024zpl,Wang:2018fyk}. The appearance of the very
low transition redshift $z_{\rm crit} \simeq 0.08$ in CPL \textit{Planck}+DESI
is the direct geometric signature of this near-saturation of the
$\Omega_{\rm DE}$--$w_0$ condition~\cite{DESI:2025zgx,Cortes:2024lgw,Wang:2025bkk,Giare:2024beyond,Giare:2025overview}.

\subsubsection{Summary of the CPL comparison}

In summary, within $w_0w_a$CDM the \textit{Planck}+DESI~BAO combination does not
unambiguously establish present day acceleration: the data do not extend close
enough to $z=0$ to break the $\Omega_{\rm DE}$--$w_0$ degeneracy, and the inferred
$q_0$ is consistent with both signs~\cite{Chevallier:2000qy,Linder:2002et,DESI:2025zgx,Cortes:2024lgw,Giare:2024beyond}. SDSS, thanks to the lower
effective redshift of the MGS, returns a clearer (though still moderate)
detection of acceleration~\cite{Ross:2014qpa,eBOSS:2020yzd}, and adding Pantheon+ to either combination
resolves the ambiguity~\cite{Scolnic:2021amr,Brout:2022vxf}. Consistently, when the MGS the lowest redshift
SDSS anchor at $z_{\rm eff}=0.15$ is removed (Sec.~\ref{subsec:noMGS}), the
lowest SDSS effective redshift rises to $z_{\rm eff}=0.38$, comparable to the
lowest DESI bin, and the SDSS $q_0$ constraint correspondingly weakens and drifts
toward the DESI value. We develop the implications of this comparison in the
Conclusions.

\section{Conclusions}\label{sec:Conclusions}

In this work we compared the late-time expansion histories inferred from the
SDSS DR16 and DESI DR2 (Year-3) BAO datasets, combined with \textit{Planck}
CMB information, within four cosmological models. As detailed in
Sec.~\ref{sec:Data}, the CMB and lensing inputs are similar but not identical
across the SDSS and DESI combinations (different likelihood versions, different
lensing datasets, and the presence of RSD in the SDSS chains); we have therefore
treated the survey-to-survey differences as directional trends rather than
calibrated significances. For the single parameter
extensions of $\Lambda$CDM ($w$CDM, $\nu$CDM, and $w$CDM\,$+\sum m_\nu$) the two
surveys yield mutually consistent constraints and a present day expansion that
remains firmly accelerating~\cite{DiValentino:2020evt,Lesgourgues:2006nd,Hannestad:2005gj}. The picture changes only in the $w_0w_a$CDM (CPL)
model, where the additional freedom along the $w_a$ direction makes the
inference markedly sensitive to the low redshift reach of each dataset~\cite{Chevallier:2000qy,Linder:2002et,Cortes:2024lgw,Giare:2024robust,Wolf:2025robustness}.

Within CPL we found that \textit{Planck}+SDSS prefers $w_0 = -0.71^{+0.19}_{-0.18}$
and a negative deceleration parameter, $q_0 = -0.22^{+0.20}_{-0.21}$, consistent
with ongoing acceleration, whereas \textit{Planck}+DESI prefers a higher
$w_0 = -0.41^{+0.21}_{-0.22}$ and a positive $q_0 = +0.10^{+0.21}_{-0.23}$,
placing the median posterior on the non-accelerating side of the $q_0 = 0$
boundary. We argued that this contrast is not driven by any single discrepant
bin, nor by genuinely new physics, but by the difference in the lowest
\emph{effective} redshift probed by each survey: $z_{\rm eff} \approx 0.15$ for
the SDSS MGS against $z_{\rm eff} \approx 0.295$ for the DESI BGS. Because $w_0$ and $q_0$ are present day quantities, the lowest available BAO
distance measurement, reported at the BGS effective redshift
$z_{\rm eff}\simeq 0.30$, provides no direct anchor closer to $z=0$. The CPL
reconstruction must therefore extrapolate the equation of state down to the
present epoch, driving $w_0$ toward zero and $q_0$ toward positive values along
the $\Omega_{\rm DE}$--$w_0$ degeneracy direction~\cite{CaldwellKamionkowski2009,DiValentino:2020evt,Cortes:2024lgw}.

Two independent tests confirm this interpretation. Adding the Pantheon+
supernova compilation to \textit{Planck}+DESI restores the missing low redshift
leverage, returning a clearly accelerating Universe ($q_0 = -0.37 \pm 0.06$) in
good agreement with SDSS and shrinking the uncertainties by nearly a factor of
four~\cite{Scolnic:2021amr,Brout:2022vxf}. Conversely, removing the low redshift MGS anchor from \textit{Planck}+SDSS
weakens its constraint and drifts $q_0$ upward toward the DESI value, landing
between the two surveys~\cite{Ross:2014qpa,eBOSS:2020yzd,DESI:2025zgx}.

We therefore conclude that the apparent DESI preference for a non-accelerating
present epoch \emph{in the BAO+CMB combination} may, at least in part, reflect its
redshift sampling rather than a true departure from $\Lambda$CDM~\cite{DESI:2024kob,Cortes:2024lgw,Giare:2025overview}. We emphasize that
this statement concerns the inference of $q_0$ (and $w_0$) and does not by itself
account for the full $w_0w_a$CDM preference obtained when supernovae are added,
which carries its own low redshift information. A robust determination of the
present dynamical state of the Universe in flexible dark energy models requires
distance information extending all the way down to $z \to 0$. We accordingly
recommend that future DESI analyses adopt finer tomographic binning of the Bright
Galaxy Survey, analogous to the treatment already applied to the LRG samples, so
as to access lower effective redshifts~\cite{DESI:2024uvr,DESI:2024mwx,DESI:2025zgx}. Since the BGS BAO sample spans $0.1<z<0.4$, even a simple two-bin split (e.g.\ $0.1<z<0.25$ and $0.25<z<0.4$) would place the lower anchor at $z_{\rm eff}\sim0.18$, directly comparable to the SDSS MGS measurement at $z_{\rm eff}=0.15$. Whether such binning yields a
useful low redshift anchor will depend on the limited cosmic volume (and hence
BAO signal-to-noise) available at $z\lesssim0.2$; a dedicated forecast,
quantifying the achievable lowest $z_{\rm eff}$ and the resulting improvement on
$w_0$ and $q_0$, is the natural next step and is left to future work. If a
low redshift BGS bin can be measured with sufficient precision, we expect it to
sharpen the determination of present day acceleration and to test directly
whether the \textit{Planck}+DESI inference moves toward the \textit{Planck}+SDSS
and $\Lambda$CDM expectations.

\section*{Acknowledgments}
We thank Jose Beltrán Jiménez and Shadab Alam for valuable discussions. 
A.M.\ and A.C.F.\ acknowledge Prof.\ Francesco Piacentini for granting access to computational resources at CINECA under the project \texttt{INF26\_lspe}, and Matias Liempi for access under the project \texttt{IsCd2\_NSCBHs}. 
R.\ acknowledges financial support from Grant PID2024-158938NB-I00, funded by MICIU/AEI/10.13039/501100011033 and co-funded by ERDF/EU (``A way of making Europe''), as well as from Project SA097P24, funded by the Junta de Castilla y León. 
R.\ further acknowledges the use of the HPC facility Pegasus at IUCAA, Pune, India.

\bibliography{main} 

\end{document}